\newcommand{\dm}{\mbox{${\rm d}M$}}

\newcommand{\dr}{\mbox{${\rm d}$}}
\newcommand{\dndt}{\mbox{${\rm d}N/{\rm d}t$}}

\newcommand{\mmin}{\mbox{$M_{\rm min}$}}
\newcommand{\mmax}{\mbox{$M_{\rm max}$}}
\newcommand{\mup}{\mbox{$M_{\rm up}$}}

\newcommand{\mmaxth}{\mbox{$M_{{\rm max, 3rd}}$}}

\newcommand{\nbin}{\mbox{$N_{\rm bin}$}}
\newcommand{\mtot}{\mbox{$M_{\rm tot}$}}
\newcommand{\msun}{\mbox{$M_{\odot}$}}

\newcommand{\pyr}{\mbox{yr$^{-1}$}}
\newcommand{\fmid}{\mbox{$f_{\rm MID}$}}
\newcommand{\tdis}{\mbox{$t_{\rm dis}$}}

\documentclass{aa}  

\usepackage{graphicx,natbib}
\bibpunct{(}{)}{;}{a}{}{,} 

\usepackage{txfonts}
\begin{document}
   \title{An alternative method to study star cluster disruption}

   \author{M. Gieles
          \inst{1}
          \and
          N. Bastian\inst{2}
          }

   \offprints{mgieles@eso.org}

   \institute{European Southern Observatory, Casilla 19001, Santiago 19, Chile\\
              \email{mgieles@eso.org}
         \and
             Department of Physics and Astronomy, University College London, Gower Street, London WC1E 6BT\\
             \email{bastian@star.ucl.ac.uk}
             }

   \date{Received October 23, 2007; February 16, 2008}

  
  \abstract{
  Many embedded star clusters do not evolve into long-lived bound clusters. The most popular explanation for this ``infant mortality" of young (few Myrs) clusters is the expulsion of natal gas by stellar winds and supernovae, which perturbs the clusters' potential and leaves up to 90\% of them unbound.   
  A cluster disruption model has recently been proposed in which this mass-independent disruption of clusters proceeds for another Gyr after gas expulsion. In this scenario, the survival chances of massive clusters are much smaller than in the traditional mass-dependent disruption models.   The most common way to study cluster disruption is to use the cluster age distribution, which, however, can be heavily affected by incompleteness.  To avoid this pitfall we introduce a new method of studying cluster disruption based on size-of-sample effects, namely the relation between the most massive cluster, \mmax, and the age range sampled.  Assuming that clusters are stochastically sampled from a power-law cluster initial mass function, with index $-2$ and that the cluster formation rate is constant, \mmax\ scales with the age range sampled, such that the slope in a log(\mmax) {\it vs.} log(age) plot is equal to unity.  
  This slope decreases if mass-independent disruption is included. For 90\% mass-independent cluster disruption per age dex, the predicted slope is zero.
 For the  solar neighbourhood, SMC, LMC, M33, and M83, based on ages and masses taken from the literature, we find slopes consistent with the expected size-of-sample correlations for the first 100~Myr, hence ruling out the 90\% mass-independent cluster disruption scenario.  For M51, however,  the increase of log(\mmax) with log(age) is slightly shallower and for the Antennae galaxies it is flat.  This simple method 
 shows that the formation and/or disruption of clusters in the Antennae must have been very different from that of the other galaxies studied here, so it should not be taken as a representative case.
  
   \keywords{galaxies: star cluster -- galaxies: evolution}
               }

   \maketitle
%

\section{Introduction}
In this study we propose an alternative method of studying the nature of cluster disruption, by using only a small number of massive clusters and a few simple assumptions. This study is motivated by the recent debate on the duration of the ``infant mortality" phase of clusters. 

The term infant mortality was coined by Lada \& Lada~(\citeyear{2003ARA&A..41...57L}, LL03), who noticed from a comparison of the number of embedded clusters per unit time (\dndt) to the \dndt\ of optically detected clusters in the solar neighbourhood, that up to 90\% of the embedded clusters do not survive the gas expulsion phase. This rapid expulsion of gas is driven by 
  stellar winds, ionisation and supernovae of early type stars. Due to the expanding gas shells  the binding energy of the initial system of stars and gas is reduced and the stars that have remained in place (assuming instantaneous gas expulsion) suddenly have velocities higher than the local escape velocity \citep{1978A&A....70...57T, 1980ApJ...235..986H, 1984ApJ...285..141L, 1997MNRAS.286..669G, 2001MNRAS.323..988G, 2005tdug.conf..629K}. Under the assumption of a constant, or mass-independent, star formation efficiency, the fraction of clusters that becomes unbound (infant mortality rate), or the fraction of mass lost from each cluster (``infant weight loss") is independent of the mass of the embedded cluster. Theory predicts that the effects of gas expulsion should be largely over within 20 Myr \citep{2006MNRAS.373..752G}.

More recently, the infant mortality scenario has also been used to explain the steep drop in \dndt\ around $\sim10-20\,$Myr of clusters in M51. Around 70\% of the clusters in M51 do not survive past 20~Myr, roughly independent of cluster mass \citep{2005A&A...431..905B, 2005A&A...441..949G}. It is noteworthy that the study of M51 clusters was based on optically detected clusters only, while the study of the solar neighbourhood  (LL03) used a comparison between young embedded and older optically visible clusters to determine the infant mortality fraction. The \dndt\ distribution of optically detected clusters in the solar neighbourhood is nearly flat for ages $\lesssim100\,$Myr\footnote{For a sample limited to clusters within 600\,pc from the sun and with masses $>100\,\msun$. When plotting \dndt\ of all clusters, as in \citet{2007AJ....133.1067W}, it declines due to incompleteness effects.} \citep{2005A&A...441..117L}, i.e. very different than that of the M51 clusters. 

\citet{2005ApJ...631L.133F} find that the \dndt\ distribution of a mass limited cluster sample of the Antennae galaxies declines roughly as $t^{-1}$ up to $\sim1\,$Gyr. They explain this decline by infant mortality, removing 90\% each age dex independent of cluster mass over the full age range of the \dndt\ distribution. Since this time-scale is two orders of magnitudes longer than the time-scale involved in the original definition of infant mortality by LL03, we will refer to this disruption model as ``mass-independent dissolution" (MID).

Based on the Magellanic Cloud Photometric Survey, \citet{2005AJ....129.2701R} presented a set of structural parameters of clusters in the Small Magellanic Cloud (SMC). Their version of the \dndt\ distribution was declining and this was interpreted by \citet{2006ApJ...650L.111C} as MID removing the same fraction (90\%) of clusters  each age dex as in the Antennae during the first 1 Gyr.

The interpretation of \citet{2006ApJ...650L.111C} is surprising for two reasons: {\it a)} it is in disagreement with earlier studies on the \dndt\ distribution of SMC clusters, who found a flat \dndt\ up to $\sim1\,$Gyr \citep{1987PASP...99..724H,2006A&A...452..179C} and {\it b)} they are the first to suggest a disruption scenario in which the life-time of star clusters during the first Gyr does not depend on their mass  or local environment. This contradicts  the existing theoretical understanding of cluster disruption.
Clusters in the SMC should  survive much longer than clusters with similar masses in the Antennae galaxies, due to the much lower tidal field strength and molecular cloud density in the SMC (e.g. \citealt{1958ApJ...127...17S, 1988IAUS..126..393W, 1997MNRAS.289..898V,    2003MNRAS.340..227B,   2006MNRAS.371..793G}). From comparisons between the age and mass distributions of clusters in different galaxies this scenario is also supported by observations (e.g.   \citealt{1985ApJ...299..211E, 1987PASP...99..724H, 2003MNRAS.338..717B, 2005A&A...429..173L}). The decline of the \dndt\  distribution of SMC clusters reported by \citet{2006ApJ...650L.111C} is  most probably the result of detection incompleteness  since they derive the slope of the \dndt\ distribution from a fit to the full cluster sample, without making a mass cut. \citet{2007ApJ...668..268G} showed that a mass limited sub-sample has a flat \dndt\ up to 1 Gyr. This is because at older ages clusters are intrinsically fainter, causing the number of observed clusters in the full (luminosity limited) sample to drop with increasing age.

The interpretation of results based on \dndt\ distributions will always be heavily dependent on how incompleteness is determined or corrected for. To remedy these shortcomings we have developed a method that can serve as an independent check of the universal MID scenario proposed by \citet{2005ApJ...631L.133F}, \citet{2006ApJ...650L.111C} and \citet{2007AJ....133.1067W} Êand the traditional disruption scenario in which the disruption time depends on mass.

Our new method is based on a few very elementary  assumptions and needs only a handful of massive clusters at various ages, i.e. well above the detection limit, avoiding problems with incompleteness of faint clusters at old ages. An additional advantage of our method is that age dating of massive clusters is more accurate than for clusters with lower masses since the stellar IMF is well populated, so stochastic fluctuations in the cluster colours due to IMF sampling are small.  The method is similar, but under different assumptions, to that used by \citet{2007MNRAS.379...34M} who used the most massive cluster as a function age to derive the star formation history of galaxies. It is fundamentally similar to the study of  \citet{2003AJ....126.1836H} who used the most massive cluster per logarithmic age bin to constrain the initial mass function of clusters.

Using the size-of-sample effect of cluster populations, in particular the most massive cluster per logarithmic time interval, we will address the following points in this paper:
\begin{itemize}
\item Is cluster disruption independent of cluster mass and environment?
\item How does the observed relation between the most cluster found and the linear time interval probed depend on cluster disruption and the cluster formation history of the galaxy?
\item What is the relative fraction of stellar mass in bound and long lived ($\gtrsim10\,$Myr) clusters?
\end{itemize}

The proposed method is meant to compliment, and provide an independent check, of existing methods, which are normally based on the observed age distribution of full cluster populations.

In \S~\ref{sec:stat} we present the statistical considerations of the size-of-sample effect and compare the expected behavior to observed cluster populations in \S~\ref{sec:data}. A discussion and our conclusions are presented in \S~\ref{sec:discussion} and \S~\ref{sec:conclusions}, respectively.


\section{The maximum cluster mass from statistical considerations: size-of-sample effect}
\label{sec:stat}

\subsection{The mass of the most massive cluster}
We assume that masses of clusters follow from random sampling of a power-law cluster initial mass function (CIMF):
\begin{equation}
\phi(M)=\frac{\dr N}{\dr M}=A\,M^{-\alpha},
\label{eq:mf}
\end{equation}
with $\alpha\simeq2$ \citep{1999ApJ...527L..81Z, 2003A&A...397..473B, 2003dhst.symp..153W, 2003MNRAS.343.1285D, 2006A&A...450..129G}. 

For $\alpha\leq2$ the total mass diverges to infinity when integrating to infinity, so a maximum mass has to be chosen, which can be interpreted as a physical maximum above which no clusters can exist. We will refer to this mass as \mup\  (i.e. the upper limit) and to the most massive cluster observed, i.e. the most massive cluster actually formed, as \mmax.

The value of \mmax\ depends on the constant $A$ in Eq.~\ref{eq:mf} and can be found by integrating $\phi(M)$ from \mmax\ to \mup\ and setting this equal to 1. For $\alpha>1$ and $\mup>>\mmax$, which is probably true for most galaxies, this results in  $A=(\alpha-1)\,M^{\,\alpha-1}_{\rm max}$. For $\alpha=2$ this reduces to $A=\mmax$. The relation between \mup\ and \mmax\ depends on the number of clusters formed ($N$) which is proportional to $A$. The larger $N$, the closer the statistically probable \mmax\ will be to \mup. We can relate \mmax\ to $N$ as

\begin{eqnarray}
N&=&\int_{\mmin}^{\mup} A\,M^{-\alpha}\,\dm \\
	        &\simeq& \frac{M^{\alpha-1}_{\rm max}}{M^{\alpha-1}_{\rm min}} , \alpha>1\label{eq:ntot1}\\
	        &\simeq& \frac{\mmax}{\mmin}, \alpha=2\label{eq:ntot2},
\end{eqnarray}
where in the last steps we againÊ assumed $\mup/\mmax>>1$.  For a constant \mmin\ we see from Eq.~\ref{eq:ntot1} that \mmax\ scales with $N$ as 

\begin{equation}
\mmax\propto N^{1/(\alpha-1)}.
\label{eq:mmaxntot}
\end{equation}
The scaling of \mmax\  with $N$ for clusters (Eq.~\ref{eq:mmaxntot}) is analogous to what \citet{2005ApJ...620L..43O} found for stars. 
Assuming a constant cluster formation rate (CFR), \dndt=constant, the number of clusters in each equal logarithmic time interval, \nbin, increases linearly with age, since 

\begin{equation}
\nbin\propto\frac{\dr N}{\dr\ln(t)}=t\,\frac{\dr N}{\dr t}
\label{eq:dndlogt}
\end{equation}
and so $\nbin(t)\propto t$.
The same holds for  $\dr N/\dr\log(t)$ apart from an additional constant $\ln(10)$.
Therefore \mmax\ in logarithmic age bins scales as

\begin{equation}
\mmax\propto N^{1/(\alpha-1)}_{\rm bin}Ê\propto t^{1/(\alpha-1)}.
\label{eq:mmax}
\end{equation}
The above derivation was based on a constant CFR and no disruption. \citet{2003AJ....126.1836H} already noticed that a power-law relation for the CFR with age (CFR$\propto t^\eta$, with $\eta$ negative for a CFR that was lower in the past) would change the relation for \mmax\ to

\begin{equation}
 \mmax\propto t^{(1+\eta)/(\alpha-1)}.
\label{eq:mmaxcfr}
 \end{equation}

The scaling of \mmax\ with age appears since \nbin\ increases in equal sized logarithim age bins. Our assumptions are fundamentally different from those of \citet{2007MNRAS.379...34M}, since  in their model  \mmax\  is set by the star formation rate and a time-scale of formation of an entire cluster population of 10\,Myr. Though they do randomly sample  \mmax\  from a probability density function, allowing some influence of the size of the sample, they on average predict a constant \mmax\ in logarithmic age bins for a constant CFR, rather than the increase predicted by Eq.~\ref{eq:mmax}.
Also, they assume a cluster disruption law and solve for the cluster formation history.  We, on the other hand, take the cluster formation rate as constant and test a theory of cluster disruption.

For the moment we consider the simplified scenario of a constant CFR and no {\it mass-dependent} disruption (Eq.~\ref{eq:mmax}). In the next section (\S~\ref{subsec:mid}) we will add the MID scenario to this. In \S~\ref{sec:discussion}  we discuss the effects of mass-dependent disruption and variations in the cluster formation history.

\subsection{The effect of mass-independent dissolution on \mmax}
\label{subsec:mid}

We consider the scenario proposed by \citet{2005ApJ...631L.133F}, \citet{2006ApJ...650L.111C} and \citet{2007AJ....133.1067W}, in which each age dex a fixed fraction of the number of clusters gets destroyed, independent of the cluster mass.  We refer to this fraction as \fmid. They argue that 90\% of the clusters are destroyed each age dex, so $\fmid=0.9$.  This reduction in number results in an expression for the remaining clusters as a function of time of the form $\dndt\propto t^\lambda$, with $\lambda= {\log(1-f_{\rm MID})}$, which is $t^{-1}$ for $\fmid=0.9$.  For \fmid=0, 0.5 and 0.8 we find $\lambda=0, -0.3$ and $-0.7$, in agreement with the values quoted by \citet{2007AJ....133.1067W}.
From this relation for \dndt\ and Eq.~\ref{eq:dndlogt} we find that  the number of clusters per logarithmic age bin, \nbin, depends on \fmid\ as
\begin{equation}
\nbin\propto\frac{\dr N}{\dr\log(t)}\propto\,t^{1+\log(1-f_{\rm MID})}.
\end{equation}
We now substitute this expression in Eq.~\ref{eq:mmaxntot} for $N$ to find the trend of \mmax\ with age when MID is included:

\begin{eqnarray}
\mmax             & \propto & t^{(1+\log(1-f_{\rm MID}))/(\alpha-1)},
\label{eq:mmaxfin}
\end{eqnarray}
which gives the result as in Eq.~\ref{eq:mmaxcfr} for no disruption ($\fmid=0$) and a constant CFR ($\eta=0$). From a comparison between Eq.~\ref{eq:mmaxcfr} and Eq.~\ref{eq:mmaxfin} it is directly visible that there is a degeneracy between a CFR that has been increasing ($\eta<0$) and MID. MID reduces the number of clusters with increasing age. Due to the mass-independent nature of this disruption model, it is impossible to distinguish between an increasing CFR and MID.

In a plot of $\log(\mmax)$ {\it vs.} $\log$(age) we thus expect  a slope
\begin{eqnarray}
{\rm slope}&=&\frac{1+\log(1-\fmid)}{\alpha-1}\,\label{eq:slopentot}\\
		&=&+1 \hspace{2.6cm} {\rm for}\,\, \fmid=0, \,\alpha=2, \label{eq:slopentot0}\\
		&=&0   \hspace{2.82cm} {\rm for}\,\,\fmid=0.9\,\,\,{\rm and\,all\,\alpha}\label{eq:slopentot1}.
\end{eqnarray}
 
\begin{figure}
   \centering
   \includegraphics[width=9cm]{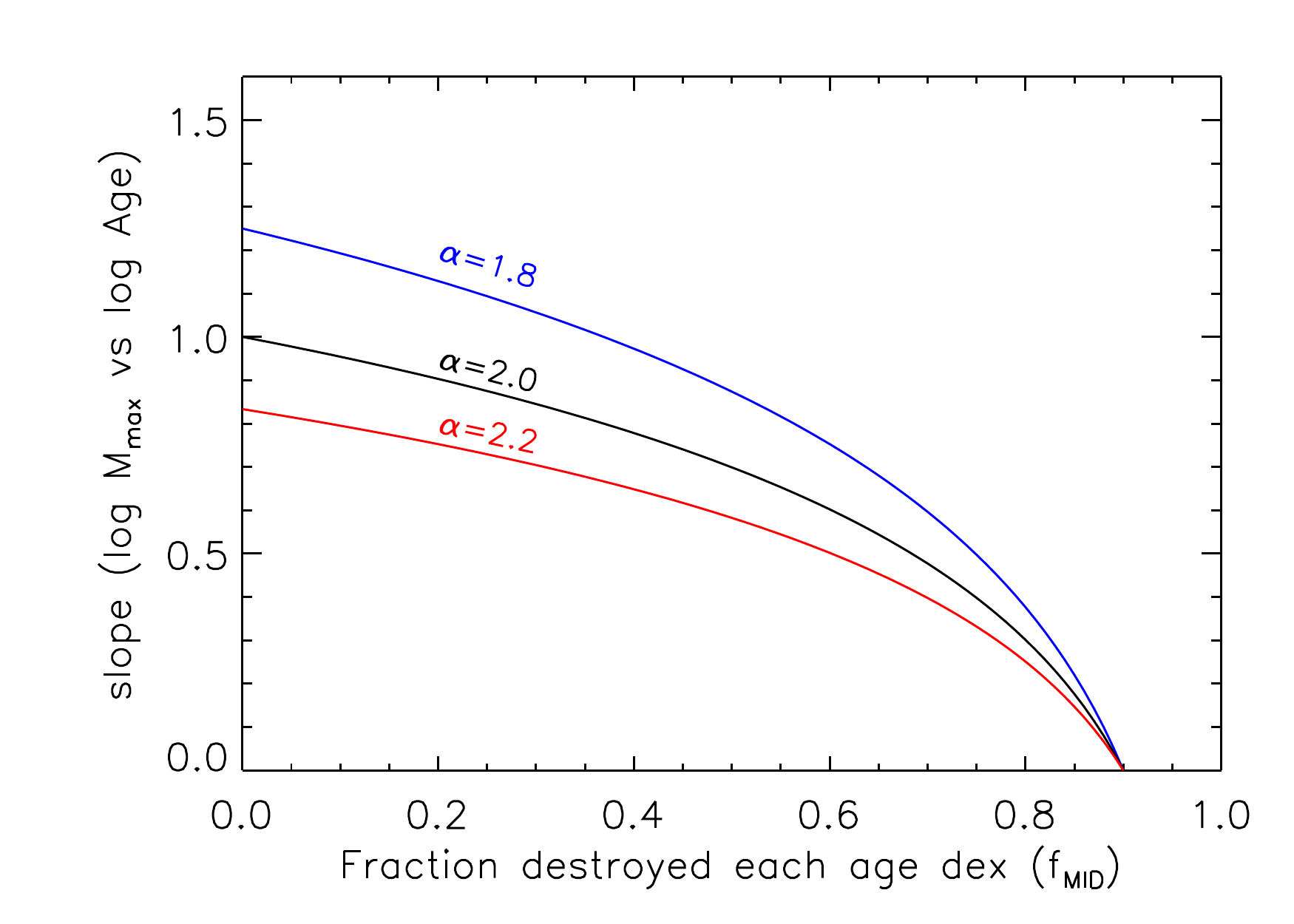}
      \caption{Predicted slopes for $\log(\mmax)$ {\it vs.} log(age) for different MID fractions (\fmid) and three different indices of the CIMF ($\alpha$) (Eq.~\ref{eq:slopentot}). }
         \label{fig:slope}
   \end{figure}

In Fig.~\ref{fig:slope} we show the predicted slopes following from Eq.~\ref{eq:slopentot} for three different values for $\alpha$. 
{\it For the universal $\fmid=0.9$ scenario, which was proposed by \citet{2007AJ....133.1067W}, the predicted slope is 0 for all values of $\alpha$.}  For lower values of \fmid, we expect noticeably different slopes than +1, providing an independent strong constraint of the acceptable values of \fmid.

%


\section{Comparison to the observations}
\label{sec:data}

\subsection{Description of the data used}
We collect cluster ages and masses in seven different galaxies from the literature:  the Milky Way (solar neighbourhood), SMC, LMC, M33, M83, M51 and the Antennae galaxies. Here we briefly mention the origin of the data and we refer the reader to these papers for details on the data reduction and age fitting techniques.

 For the clusters in the solar neighbourhood we use ages from the catalogue of \citet{2005A&A...438.1163K} and corresponding masses derived and kindly provided to us by \citet{2005A&A...441..117L}. The mass estimates are derived from the number of stars with high membership probability and an extrapolation of the stellar initial mass function.  We limit ourselves to the 209 clusters in the catalogue that are within a distance of 1\,kpc from the sun, for which the mass estimates are believed to be accurate and the sample is not severely affected by distance incompleteness. 

For the SMC and the LMC we use the results of \citet{2003AJ....126.1836H}, who kindly provided us with a table of ages and luminosities of  191 SMC clusters and 748 LMC clusters. We derived the masses of the clusters using the age-dependent mass-to-light ratios of the GALEV models \citep{2002A&A...392....1S, 2003A&A...401.1063A} with $Z=0.004$ for the SMC and $Z=0.008$ for the LMC.
For NGC~5236 (M83) we used the ages and masses derived by Mora et al. (2008, in prep), who kindly provided us with a table containing ages and masses of 219 clusters.
For M33 we used a recent catalogue of 201 clusters published by \citet{2007AJ....134..447S}.
The M51 data were taken from \citet{2005A&A...431..905B}.
We derive the ages and masses of Antennae clusters from Fig.~2  of \citet{1999ApJ...527L..81Z},  and converted luminosities to masses using the Bruzual \& Charlot (1996, unpublished) SSP models, which were used by \citet{1999ApJ...527L..81Z} to derive the ages.

\subsection{Observed trends of $\log(\mmax)\,\,vs. \log$(age)}

   \begin{figure*}
   \centering
   \includegraphics[width=17.5cm]{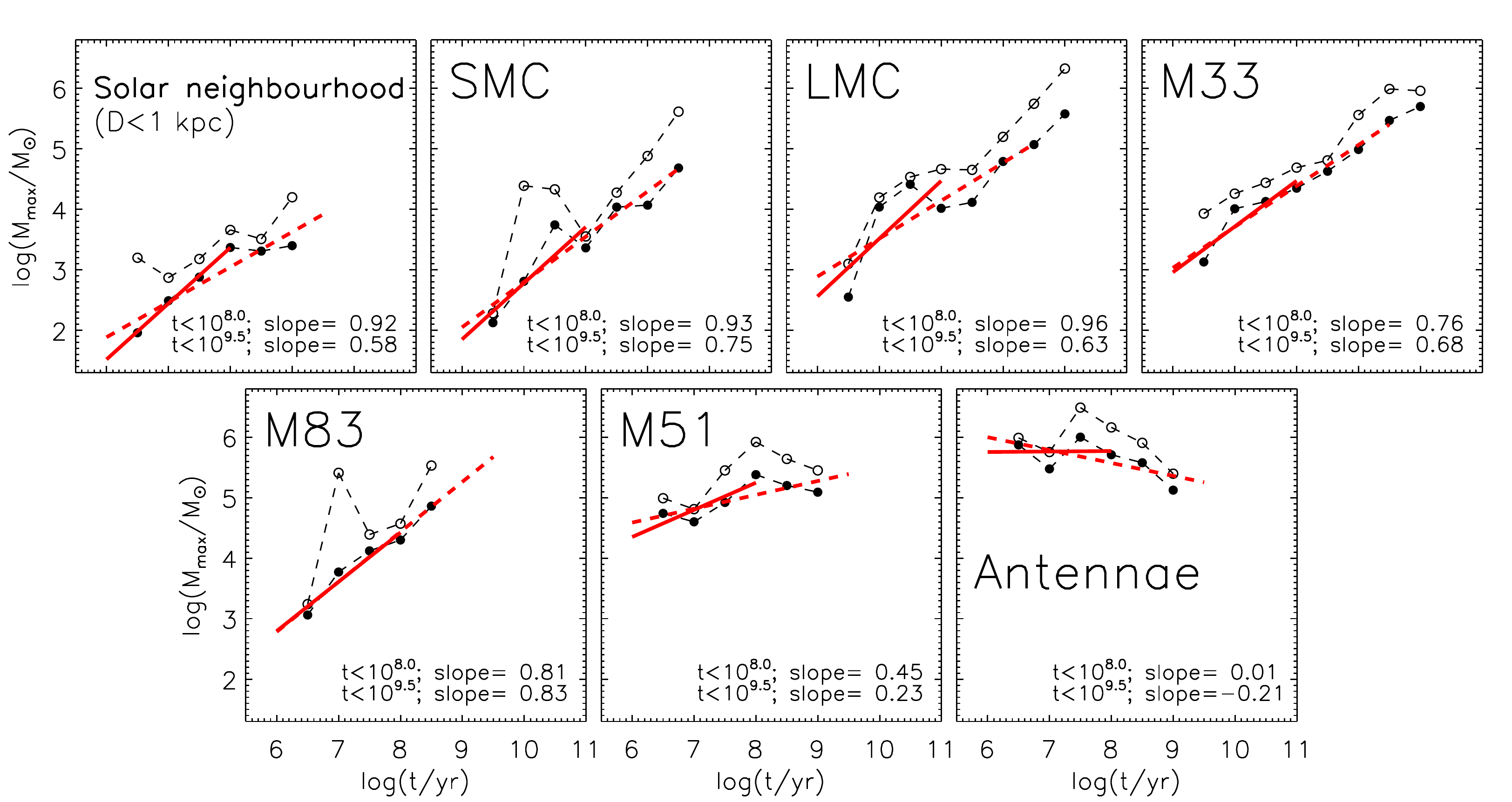}
   \caption{Evolution of \mmax\ with equal size log(age) bins for the seven galaxies in our sample (open circles). The third most massive clusters (\mmaxth) in each bin are shown as filled circles. Fits to $\log(\mmaxth)$ {\it vs.} log(age) on two different age ranges are shown as dashed and full lines.}
   \label{fig:max}
    \end{figure*}
In Fig.~\ref{fig:max} we show \mmax\ as a function of log(age) in bins of 0.5 dex as open circles. We also show the third most massive cluster in each age bin (\mmaxth) as filled circles,  which is expected to follow the same relations as the most massive, though with less scatter. We note that the trends for the second, fourth and fifth most massive cluster are all similar to the trend of \mmaxth. The full lines represent the result of fits of $\log(\mmaxth)$ with log(age) over the first $100\,$Myr and the dashed lines consider a time range of 3 Gyr.  The scatter of the data points around the fit is most likely due to stochastic effects and should not be interpreted as variations in the CFR. The probability density function (PDF) of the most massive object drawn from a power-law distribution has an asymmetric  spread around the mean, but in logarithmic units the shape and width PDF is independent of the value of \mmax\ \citep{2002AJ....124.1393L, 2007MNRAS.379...34M}. Stochastic effects will therefore not affect our results.

For the solar neighbourhood, SMC, LMC, M33 and M83 the observed increase of $\log(\mmaxth)$ in the first $100\,$Myr is consistent with the size of sample prediction without disruption (Eqs.~\ref{eq:mmax}\&\ref{eq:slopentot0}). The slopes are not exactly $+1$, but we can constrain \fmid\ using Eq.~\ref{eq:slopentot} to $\fmid<0.2$ for the solar neighbourhood, SMC and LMC and $\fmid<0.4$ for M33 and M83. {\it This rules out the long term 90\% ($\fmid=0.9$) MID proposed by \citet{2007AJ....133.1067W}  as a universal cluster disruption mechanism for these galaxies. For this to be true, \mmax\ had to be constant with log(age) (Eq.~\ref{eq:slopentot1}) for all galaxies.}

The trends for M51 and the Antennae are clearly different from the other five galaxies. The slow or lack of increase of $\log(\mmax)$ for M51 in the first $\sim10-20\,$Myr is consistent with the $70\%$ infant mortality, independent of mass, over that time range as determined by \citet{2005A&A...431..905B}.  
After log(age)=7 the value of  \mmax\ is increasing again, confirming that the infant mortality phase lasts for about 10\,Myr only.
 The flat slope beyond $100\,$Myr in M51 was interpreted as  a truncation of the CIMF  around $\mup\simeq5\times10^5\,\msun$ \citep{2006A&A...450..129G, 2006A&A...446L...9G}. Note that these three phases can not  be derived from the trend of \mmax\ with log(age) only, since we have only five data points. The trend does support the results derived from the \dndt\ distribution and the luminosity function by  \citet{2005A&A...431..905B} and  \citet{2006A&A...450..129G, 2006A&A...446L...9G}
 
 The flat relation for the Antennae galaxies is consistent with $90\%$ disruption each age dex (\fmid=0.9) during a Gyr. Note, however,  that if this result is attributed entirely to \fmid=0.9 \citep{2007AJ....133.1067W}, which implies that there is no truncation of the CIMF and that the CFR has been constant over this age range, which was proposed by  \citet{2007AJ....133.1067W}, then based on size-of-sample effects we would expect that the galaxy was producing clusters with masses up to $10^9\,\msun$  in the oldest log(age) bin, but they have been destroyed due to MID. This seems unlikely given what we know about other galactic mergers. Namely, the major burst of star formation in the Antennae still has to happen,which will be when the nuclei coalesce (e.g. \citealt{2006MNRAS.373.1013C}) and there are no star cluster known with masses in excess of $10^8\,\msun$.

\subsection{Deriving the formation rate in bound clusters}
\label{sec:cfr}
Under the assumption of a constant formation rate, a power-law CIMF with index $-2$ and no disruption of clusters, 
the values of \mmax\ are indicative of the amount of stellar mass formed in bound clusters. Here we will use the SMC as an example, since it has been shown that for the clusters in this galaxy the disruption time-scale is long \citep{1987PASP...99..724H, 2003MNRAS.338..717B, 2007ApJ...668..268G}.  In addition, this is one of the only galaxies for which a global star formation rate (SFR) has been determined and the full galaxy has been imaged for its cluster population. This is necessary since \mmax\ scales with the total number of observed clusters and therefore with the fraction of the galaxy that has been imaged. 

For $\alpha=2$ the total mass formed in clusters, \mtot,  can be found from

\begin{eqnarray}
\mtot&=&\int_{\mmin}^{\mup} \,M\,\phi(M)\,\dm\\
	&\simeq&\mmax\,\ln\left(\frac{\mup}{\mmin}\right) \label{eq:mtoteqtwo}\\
	&\simeq&10\,\mmax \label{eq:mtoteqtwo2},
\end{eqnarray}
where we again assume $\mup>>\mmax$ and $\mup/\mmin=10^5$, e.g. $\mmin=10^2\,\msun$ and $\mup=10^7\,\msun$. For $\alpha=2$ the predicted \mmaxth\ is simply one third of \mmax.

When representing the data as in Fig.~\ref{fig:max}, we need to assume a CFR, i.e. the amount of mass formed in bound clusters per unit of time. With this and the width of the age bins we can calculate $\mtot=$CFR$\times\Delta t$. 
We find good agreement between the SMC data in Fig.~\ref{fig:max} and a prediction for \mmax\ and \mmaxth\ with CFR$\simeq2\times10^{-3}\msun\,\pyr$. Due to the scatter in the data this is only a rough prediction, accurate to within a factor of two.

We can compare the prediction of the CFR to the star formation rate (SFR). \citet{2004AJ....127.1531H} derive a mean global star formation rate of $0.1\,\msun\pyr$ based on the field star population and \citet{2004A&A...414...69W} derive SFR=$0.05\,\msun\pyr$ based on far-infrared observations of the SMC. Our derived CFR of $2\times10^{-3}\,\msun\pyr$ represents only $2-4\,$\% of the total SFR. {\it This implies that  $\sim$97\% of the star formation occurs in a dispersed fashion, or, that the infant mortality rate in the SMC is close to 97\%.} This data does not allow to tell these two scenarios apart.

For the solar neighbourhood, a similar ratio of CFR/SFR$\simeq0.05$ was found \citep{2006A&A...455L..17L, 2007astro.ph..2166L} from a comparison of the star formation rate in (optical detected) clusters relative to the star formation in field stars, consistent with the infant mortality rate derived by LL03.

\section{Discussion}
\label{sec:discussion}
\subsection{Is there no infant mortality of clusters in some galaxies? }
We conclude that the slopes for the first $100\,$Myr in the solar neighbourhood, SMC, LMC, M33 and M83 are consistent with a complete lack of mass-independent disruption. This implies that the \dndt\ of {\it mass limited} cluster samples should be flat over this age range and that a {\it luminosity limited} cluster sample should follow the decline predicted by the fading of clusters, i.e. $\dndt\propto t^{-\zeta}$, with $-1.0<\zeta<-0.7$ \citep{2003MNRAS.338..717B, 2007ApJ...668..268G}. For the SMC this was recently shown to be the case by \citet{2007ApJ...668..268G} and \citet{ 2008MNRAS.383.1000D}. The \dndt\ of the LMC clusters also follows the fading prediction in the first $100\,$Myr \citep{2006MNRAS.366..295D}.  \citet{2008MNRAS.383.1103P} show that the \dndt\ of massive clusters  is nearly flat. Both these findings and the near linear increase of log(\mmax) with log(age) we see in  Fig.~\ref{fig:max} suggest that there can not be 90\% MID of clusters in the LMC.

The \dndt\ of M33 clusters declines as $t^{-1.1\pm0.1}$ \citep{2007AJ....134..447S}, which was interpreted by the authors as rapid dissolution.  However, the authors  also note that their sample is luminosity limited. When the sample is limited by a detection in a blue filter, such as $B$ or $U$, the \dndt\ distribution of a sample that is not affected by any disruption declines as $t^{-1}$.
This means that both the \dndt\ distribution as shown by \citet{2007AJ....134..447S} and the nearly linear increase of $\log(\mmax)$ with log(age) (Fig.~\ref{fig:max}) are in agreement and imply no mass independent disruption (i.e.  \fmid=0) in the first $100\,$Myr in M33.

 The flat \dndt\ distributions of mass-limited cluster samples  and the results presented in this work are not to be interpreted as no infant mortality of clusters in these galaxies. This is because all of the cluster populations used in this study were optically selected, i.e. no embedded clusters were included.  In fact, our findings are in perfect agreement with the 90\% infant mortality scenario for the solar neighbourhood by LL03, since they derived this from a comparison between embedded clusters to open clusters (see \S~\ref{sec:cfr}).  Our results suggest that for some galaxies the infant mortality of clusters can not be observed from optically selected clusters. The cluster samples of M51 and the Antennae are the only ones for which it has been shown that there is a steep drop in \dndt\ around $\sim10\,$Myr in a mass limited, optically selected sample.

\subsection{What happens after 100 Myr?}
 
The slopes  determined over 3 Gyr (Fig.~\ref{fig:max}) are shallower for most samples, but it is difficult, if not impossible, to come up with an explanation where MID disruption starts after $100\,$Myr.  There are several culprits for the observed flattening: stellar evolution combined with ``standard" mass-dependent disruption, a CIMF that is steeper or truncated at high masses ($\sim10^6\,\msun$) or a non-constant CFR.  Below we will discuss each of these effects in turn.

\subsubsection{The effect of a non-constant cluster formation history}
\label{ssec:cfr}
 So far we have assumed, for simplicity, that the star/cluster formation rate has been constant in time. For the age range of $10-100\,$Myr this is probably a reasonable assumption for most of the galaxies considered in this study. However, for the age range of 3\,Gyr, as considered in Fig.~\ref{fig:max}, this assumption may not hold and can lead to a confusion between cluster disruption and variations in the  formation history. For example, for the SMC it is generally believed that the assumption of a constant CFR is not too bad. However, when approximating the result of \citet{2004AJ....127.1531H}  for the global SFR($t$) as $t^{\eta}$, we find $\eta=-0.15$. Assuming that the CFR follows the SFR, this partially explains the  small difference between the observed slope of $+0.75$ (Fig.~\ref{fig:max}) and the predicted slope of $+1$ (Eq.~\ref{eq:mmax}), since including the effect of this increasing CFR($t$) predicts a slope of $+0.85$ (Eq.~\ref{eq:mmaxcfr}).

For the Antennae galaxies the assumption of a constant CFR over the past Gyr \citep{2005ApJ...631L.133F, 2007AJ....133.1067W} is less likely. According to the numerical models of \citet{1988ApJ...331..699B} the first encounter between NGC~4038 and NGC~4039 was around $\sim200\,$Myr ago and the models of \citet{1993ApJ...418...82M} predicted that the SFR has been increasing by at least a factor of five since then. The $t^{-1}$ disruption model introduced by \citet{2005ApJ...631L.133F} was derived from only 3 or 4 histogram points of the \dndt\ of the Antennae clusters. A combination of 90\% infant mortality in the first 10-30\,Myr and a CFR that started increasing a few 100 Myrs ago could also very well explain the trend in Fig.~\ref{fig:max} {\it and} the steep decline in \dndt\ reported by \citet{2005ApJ...631L.133F}. 

 There will always be a natural bias to studies of star cluster populations in galaxies which are actively forming stars and clusters at the present day, such as the Antennae galaxies, simply because they are more attractive to study. The degeneracy in this method between changes in the CFR and disruption and/or a truncation in the CIMF is hard to quantify.  A note of caution should be placed that the assumption of a constant CFR can lead to false detections of MID. This is not only the case when studying \dndt, but as well as \mmax\ {it vs.} log(age).

\subsubsection{Mass dependent disruption}
 To flatten the relation between $\log(\mmax)$ and $\log($age$)$ by {\it mass dependent disruption}, the dependence of the disruption time ($\tdis$) on mass has to be weaker than linear, i.e. $\tdis\propto M^{\gamma}$ where $\gamma < 1$.  When $\tdis\propto M$ and \mmax\ scales also linear with age due to size of sample effects, than all clusters lose the same fraction of their initial mass, which would not flatten the relation. From observations and theory it is expected that \tdis\ scales as $M^{0.6}$ \citep{2005A&A...429..173L, 2006MNRAS.371..793G}, from which a flattening of $\mmax$ with age is expected. \citet{2006A&A...450..129G} showed that the trends of \mmax\ with age for the SMC and LMC can be explained by mass loss at old ages due to stellar evolution and tidal effects.

\subsubsection{The cluster initial mass function}
Since the measured slope in Fig.~\ref{fig:max}  is a combination of the index of the CIMF and disruption (Eq.~\ref{eq:slopentot}), the shallower slopes could also be explained by a steeper CIMF.

We note that a slope of +0.75 in Fig.~\ref{fig:max} follows from  $\alpha\simeq2.35$ (Eq.~\ref{eq:mmax}, \citealt{2003AJ....126.1836H}.). The same high value, i.e. larger than $2$, for the index of the CIMF was found from the analyses of young ($<10\,$Myr) clusters through the increase of \mmax\ as a function of number of clusters in different galaxies \citep{2004MNRAS.350.1503W}. A value of $-2.3$ was also found for the index of the cluster luminosity function through the increase of the most luminous cluster as a function of the number of clusters \citep{2002AJ....124.1393L, 2003dhst.symp..153W, 2006A&A...450..129G}. 

Since these studies consider young clusters, i.e. not affected by disruption yet, a Schechter type upper mass limit to the CIMF is a logical scenario that can also explain the slopes slightly shallower than $+1$ in Fig.~\ref{fig:max} for the fits on the 3\,Gyr age range. Random sampling from a CIMF that is a power-law with an exponential cut-off around $M_*=10^6\,\msun$ \citep{2006A&A...450..129G, 2006A&A...446L...9G}  can result in a slope of $\sim0.75$  in the $\log(\mmax)$ {\it vs} age relation.  When sampling cluster masses from such a CIMF, the relation of \mmax\ with $N$ still resembles a straight line, i.e. the cut-off is not detectable as such as long as $\mmax\lesssim M_*$. In this scenario slightly steeper slopes at young ages are expected, since there the masses are well below the cut-off mass. This can also explain the nearly flat slopes of M51 and the Antennae, since the mass of the most massive cluster is already close to  $M_*$ at young ages and can therefore not increase more with age.  We refer to 
 \citet{2008arXiv0801.2676G} for a more detailed discussion on how a physical maximum to the cluster mass  can be derived using the method presented in \S~\ref{sec:stat}.


\section{Conclusions}
\label{sec:conclusions}
We have studied the evolution of the maximum cluster mass ($\mmax$) in bins constant in log(age) (i.e. increasing width in linear age) in different galaxies. Under the assumption of a constant cluster formation rate, a power-law cluster initial mass function (CIMF) with index $-2$, and no mass loss of clusters due to disruptive effects, we predict a linear increase of $\log(\mmax)$ with log(age), a slope of +1, when sampling the cluster masses stochastically from the CIMF.  Including the effects of mass independent disruption (MID) causes the slope to decrease in proportion to the fraction of clusters removed per age dex, \fmid\ (Eq.~12).  This results in a flat relation between $\log(\mmax)$ with log(age) when $\fmid=0.9$, the value reported by \citet{2005ApJ...631L.133F}, \citet{2006ApJ...650L.111C} and \citet{2007AJ....133.1067W}.

Based on a comparison with observed cluster populations in seven galaxies we conclude that:

\begin{enumerate}
\item The predicted linear increase of $\log(\mmax)$ with log(age) for a power-law CIMF with index $-2$ and no disruption is observed in the first $100\,$Myr for clusters in the solar neighbourhood, SMC, LMC, M33 and M83. This implies that no significant disruption has taken place (after the clusters have left the embedded phase) and with this we can rule out the scenario in which 90\% of the clusters are destroyed each age dex, independent of the cluster mass. Comparing the observed slopes with those predicted for different mass independent disruption fractions, we can constrain $\fmid < 0.2$ for the solar neighbourhood, SMC and LMC and $\fmid<0.4$ for M33 and M83.
\item For M51 the observed slope is slightly shallower, consistent with the observations of \citet{2005A&A...431..905B} who found that 70\% of the clusters in M51 do not survive the first $10-20$~Myr.   If we only look at ages between 20 and 100~Myr the slope is steeper which agrees well with no mass-independent disruption, providing additional evidence that this mass independent disruption phase lasts only 10-20~Myr and is due to the expulsion of gas.
\item In the Antennae galaxies $\log(\mmax)$ is flat in the first $100\,$Myr and even decreasing for the full age range. These observations are consistent with the scenario of 90\% MID each age dex. We note, however, that a non-constant CFR or a truncation of the cluster mass function at the high mass end can produce the same effect.
\item The SMC clusters show the best agreement with a prediction without disruption when assuming a CFR of $2\times10^{-3}\,\msun\pyr$. This is only 2 or 4\% of the global star formation rate in this galaxy, suggesting that the SMC is extremely inefficient in producing bound star clusters, but once they are formed they are very stable against disruption.
\end{enumerate}

\begin{acknowledgements}
We thank an anonymous referee for constructive comments.
We wish to thank Deidre Hunter, Henny Lamers and Marcelo Mora for providing their cluster measurements in electronic form.   Bruce Elmegreen, Iraklis Konstantopoulos, Henny Lamers, Simon Goodwin, Sally Oey and S{\o}ren Larsen are acknowledged for interesting discussions and comments on the manuscript. Additionally, we would like to thank the organisers of the workshop {\it Young Massive Star Clusters: Initial Conditions and Environments}, Enrique P\'{e}rez and Richard de Grijs, which provided  
an excellent forum to discuss this issue.
\end{acknowledgements}

\bibliographystyle{aa}

\end{document}